\documentstyle[11pt,epsfig]{article}
\newcommand{\ba}{\begin{array}}
\newcommand{\ea}{\end{array}}
\newcommand{\bd}{\begin{displaymath}}
\newcommand{\ed}{\end{displaymath}}
\newcommand{\be}{\begin{equation}}
\newcommand{\ee}{\end{equation}}
\newcommand{\bea}{\begin{eqnarray}}
\newcommand{\eea}{\end{eqnarray}}
\newcommand{\Dir}{\kern -6.4pt\Big{/}}
\newcommand{\Dirin}{\kern -10.4pt\Big{/}\kern 4.4pt}
\newcommand{\DDir}{\kern -10.6pt\Big{/}}
\newcommand{\DGir}{\kern -6.0pt\Big{/}}
\begin{document}
\
\def\bra{\langle}
\def\ket{\rangle}

\def\a{\alpha}
\def\as {\alpha_s}
\def\b{\beta}
\def\d{\delta}
\def\e{\epsilon}
\def\ve{\varepsilon}
\def\l{\lambda}
\def\m{\mu}
\def\n{\nu}
\def\G{\Gamma}
\def\D{\Delta}
\def\L{\Lambda}
\def\s{\sigma}
\def\p{\pi}

\def\etal{ {\em et al.}}
\def\mzs {M_Z^2}
\def\mws {M_W^2}
\def\q2 {q^2}
\def\sz {\sin^2\theta_W}
\def\cz {\cos^2\theta_W}
\def\lp{\lambda^{\prime}}
\def\lps{\lambda^{\prime *}}
\def\lpp{\lambda^{\prime\prime}}
\def\lpps{\lambda^{\prime\prime * }}

\def\bapp{b_1^{\prime\prime}}
\def\bbpp{b_2^{\prime\prime}}
\def\bcp{b_3^{\prime}}
\def\bdp{b_4^{\prime}}
\def\t {\times }
\def\slash {\!\!\!\!\!\!/}
\def\photino {\tilde\gamma}
\def\sel {\tilde{e}}
 \def\N10{\widetilde \chi_1^0}
                         \def\C1p{\widetilde \chi_1^+}
                         \def\C1m{\widetilde \chi_1^-}
                         \def\C1pm{\widetilde \chi_1^\pm}
 \def\Ntwo{\widetilde \chi_2^0}
                         \def\Ctwo{\widetilde \chi_2^\pm}
\def\lslep {{\tilde e}_L}
\def\rslep {{\tilde e}_R}
\def\sneu {\tilde \nu}
\def\msneu {M_\tilde \nu}
\def\mrslep {m_{\rslep}}
\def\mlslep {m_{\lslep}}
\def\mneu {m_{\neu}}
\def\mpT{p_T \hspace{-1em}/\;\:}
\def\mET{E_T \hspace{-1.1em}/\;\:}
\def\mE{E \hspace{-.7em}/\;\:}
\def\go{\rightarrow}
\def\beq{\begin{eqnarray}}
\def\Rp{R\!\!\!\!/}
\def\wrp {{\cal W}_{R\!\!\!\!/}}
\def\enq{\end{eqnarray}}
\def\goes{\longrightarrow}
\def\lsim{\:\raisebox{-0.5ex}{$\stackrel{\textstyle<}{\sim}$}\:}
\def\gsim{\:\raisebox{-0.5ex}{$\stackrel{\textstyle>}{\sim}$}\:}
 \begin{flushright}
{\large CERN-TH/2002-137}\\
{\large IPPP/02/30}\\
{\large DCPT/02/60}\\
{\large June 2002}\\
\end{flushright}
\begin{flushleft}
\end{flushleft}
\begin{center}
{\Large\bf Detection of 
heavy charged Higgs bosons}\\[5mm]
{\Large\bf at future Linear Colliders}\\[15mm]
{\Large S. Moretti\footnote{stefano.moretti@cern.ch}}\\[4mm]
{\em CERN Theory Division, CH-1211 Geneva 23, Switzerland\\
{\rm and}\\
Institute for Particle Physics Phenomenology,
University of Durham, Durham DH1 3LE, UK}
\\[20mm]
\end{center}
\begin{abstract}
\noindent
We show how a statistically significant signal 
of heavy charged Higgs bosons of a general Two-Higgs Doublet
Model produced 
in association with tau-neutrino pairs 
can be established at future Linear Colliders
in the $H^+\to t\bar b\to 4$ jet decay channel. This signature is particularly
relevant in the kinematic configuration $\sqrt s\lsim2 M_{H^\pm}$,
when the pair production channel $e^+e^-\to H^-H^+$ is no
longer available. Here, the initially overwhelming background, constituted
by top quark pair production and decay, can vigorously be reduced thanks
to a dedicated selection procedure that allows one to extract a signal
in a region of several tens of GeV around $M_{H^\pm}\approx \sqrt s/2$,
for $\tan\beta\gsim40$.
  
\end{abstract}

\vskip 1 true cm


\newpage
\noindent
Charged Higgs bosons, $H^\pm$,
 appear in the particle spectrum of a general Two-Higgs Doublet
Model (2HDM). We are concerned here with the case of Type II THDMs. In
this context, the importance of singly produced charged Higgs bosons
at future Linear Colliders
(LCs) \cite{eebiblio} has been emphasised lately in several instances
\cite{singleHpm,proceedings,Lorenzo,Logan,WH}. To begin with, it should
be recalled that the detection of $H^\pm$ states would represent an
unequivocal evidence of physics beyond the Standard Model (SM). 
Moreover, one may well face the following
situation, as a result of the Large Hadron Collider (LHC) runs: only one light 
(below 130 GeV or so) neutral Higgs boson is
found, $h$, and this is degenerate with the SM Higgs state.
For example, this can happen over a large portion of the Minimal 
Supersymmetric Standard Model (MSSM) parameter space, in the so-called
`decoupling-limit', namely, when $M_A\gg M_h$ and for intermediate 
to high values
of $\tan\beta$ (the two parameters that entirely define the Higgs
sector of a 2HDM at tree-level). This also implies that the
other MSSM Higgs states, $H$ and $H^\pm$, are similarly heavy
(i.e., $M_A\approx M_H\approx M_{H^\pm}$). 
Under these circumstances, one may have to wait for the advent of LCs, 
where precision tests of the Higgs sector can be performed, 
in order to fully clarify the nature of the Electroweak Symmetry Breaking
(EWSB) mechanism. If the existence of a (light) Higgs state will have
been proven at the LHC, then the most likely schedule at a LC will be to 
start running at a rather low energy (say, $\sqrt s=350$ or 500 GeV), 
where the corresponding Higgs 
production cross section (via $e^+e^-\to Z^*\to Zh$) is largest,
as the latter proceeds via $s$-channel annihilation. At such energies,
the heavier Higgs states may not be produced in the usual pair 
production channels
\cite{Higgs}, either because below threshold (i.e., $M_A+M_H, M_{H^+}+
M_{H^-}> \sqrt s$) or since the intervening MSSM coupling in the decoupling 
limit becomes zero (e.g., in the $ZZH$ vertex). Whereas in the neutral Higgs
sector the heavy $H$ and $A$ resonances can always be accessed in $\gamma
\gamma$ collisions  (via $\gamma\gamma\to$
`triangle loop' $\to H/A$), this is not possible for the charged
Higgs boson states, because of electromagnetic (EM) charge conservation.
In the large $\tan\beta$ region, for neutral Higgs
states, one could 
alternatively resort to the associate production mode $e^+e^-\to b\bar b
H/A$. The corresponding channel for a charged Higgs boson would be
$e^+e^-\to b\bar t H^+$, which has an additional large mass in the final
state (i.e., $m_t=175$ GeV).  

Hence, it becomes clear the importance of also studying
production modes of charged Higgs 
bosons with only one such particles in 
the final state, in order to cover the mass range $M_{H^\pm}\gsim\sqrt s/2$
at future LCs,
where $e^+e^-\to H^-H^+$ \cite{eeHpHm} falls short of a detectable cross
section \cite{Higgs}.  An analysis of the various single production modes
was performed in Ref.~\cite{singleHpm}, limitedly to their inclusive
rates. 
(For similar studies of charged Higgs bosons produced
in single modes in the case of $e\gamma$ and $\gamma\gamma$
collisions, see Refs.~\cite{egamma,2gamma}.)
There, it was shown that
only two channels offer some chances of detection:
\begin{eqnarray}
e^+e^- &\to& \tau^-\bar\nu_\tau H^+, \tau^+\nu_\tau H^-~~
\mathrm{(tree\ level)},
\label{proc_tau} \\
e^+e^- &\to& W^\mp H^\pm ~~\mathrm{(one\ loop)}.    \label{proc_wh}
\end{eqnarray}
The former is relevant in the large $\tan\beta$ region, whereas the
latter is important for the low one. As LEP2 data seem to prefer large values 
of $\tan\beta$, at least in the MSSM \cite{LEPTRE}, we attempt here to
devise a selection procedure that may help to extract process 
(\ref{proc_tau}) from the irreducible background\footnote{We defer
a similar study of channel (\ref{proc_wh}) to a forthcoming
publication.}. Since, as shown in
\cite{singleHpm}, the production rates of process (\ref{proc_tau})
are rather small in general over the mass region $M_{H^\pm}\gsim\sqrt s/2$ 
(for sake of illustration, we adopt here $\sqrt s=500$ and 1000 GeV), it
is mandatory to resort to the main decay channel of heavy charged Higgs
bosons, i.e., $H^+\to t\bar b$. Hence, the following processes are of
relevance for the signal ($S$) and the irreducible background
($B$)\footnote{Charged 
conjugated (c.c.) channels are assumed too throughout the paper.}:
\begin{eqnarray}\label{signal}
e^+ e^- &\rightarrow& \tau^- \bar\nu_\tau H^+ \to 
\tau^- \bar\nu_\tau t\bar b~~~({\mathrm{signal}}),\\ 
\label{background}
e^+ e^- &\rightarrow& \bar t t\to \tau^- \bar\nu_\tau t\bar b  
~~~~~({\mathrm{background}}).
\end{eqnarray}  
We  require the emerging top to decay fully 
hadronically, i.e., $t\to bW^+\to
3$ jets, whereas 
we have assumed $\tau$'s  to be tagged as narrow jets in their 
`one-prong' hadronic decays:
\begin{eqnarray}\nonumber
\tau^\pm &\rightarrow& \pi^\pm \nu_\tau~~~~~~~~~~~~~~~(12\%), \\ \nonumber
\tau^\pm &\rightarrow& \rho^\pm(\to \pi^\pm\pi^0) \nu_\tau~~(26\%),\\ \nonumber
\tau^\pm &\rightarrow& a_1^\pm (\pi^\pm\pi^0\pi^0)\nu_\tau~~~(8\%).
\end{eqnarray}
Hence, the complete signature is: 
\begin{equation}\label{signature}
\tau-{\rm{jet}}~+~p_T^{\mathrm {miss}}~+~4j.
\end{equation}

The simulation has been carried out at parton level. 
We have assumed a Type II 2HDM throughout with $\tan\beta=40$  and
$M_{H^\pm}$ ranging between 160 and 660 GeV. For the signal, we
have used  the formulae of \cite{singleHpm} 
for the production process and the program described in
\cite{BRs} for the decay rates. For the backgrounds we have used
the same code of Refs.~\cite{tt,bbWW}, also including non-$\bar t t$
contributions, in which the two
$W^\pm$ have been decayed appropriately\footnote{Also, we have
verified that the background due to $e^+e^-\to b\bar b ZZ$ \cite{bbZZ},
with one $Z$ decaying hadronically and the other into
two $\tau$'s, one of which escaping detection, is
negligible. Similarly, for the cases $e^+e^-\to b\bar b
W^\pm H^\mp$ and $e^+e^-\to b\bar b H^+H^-$ \cite{bbCC}, in which one boson
decays to tau-nu and the other to light-quark pairs,
and $W^-+4j$, with $W^-\to\tau^-\bar\nu_\tau$, computed in \cite{W4j}.}. 
All unstable particles
entering the two processes ($t, H^\pm$ and $W^\pm$) were finally generated
off-shell (i.e.,
 with the correct width). The integration over the final states 
has been performed numerically with the aid of VEGAS \cite{VEGAS}
and Metropolis \cite{Metro}. Finite calorimeter resolution
has been emulated through a Gaussian smearing in transverse momentum,
$p_T$, with $(\sigma(p_T)/p_T)^2=(0.60/\sqrt{p_T})^2 +(0.04)^2$
for all jets. The corresponding missing transverse momentum, 
$p_T^{\mathrm{miss}}$, was reconstructed from the vector sum of 
the visible jet  momenta after resolution smearing. A double tagging
of $b$-jets in the final state is implied.  
The non-running $b$-quark mass adopted for both the kinematics and the 
Yukawa couplings was $m_b=4.25$ GeV.  (The $\tau$-lepton was assumed to be 
massless throughout.) We neglect Initial State Radiation (ISR)
and beamstrahlung effects, as we expect these to have a marginal
impact on the relative behaviour of signal and background.
 
\begin{figure}
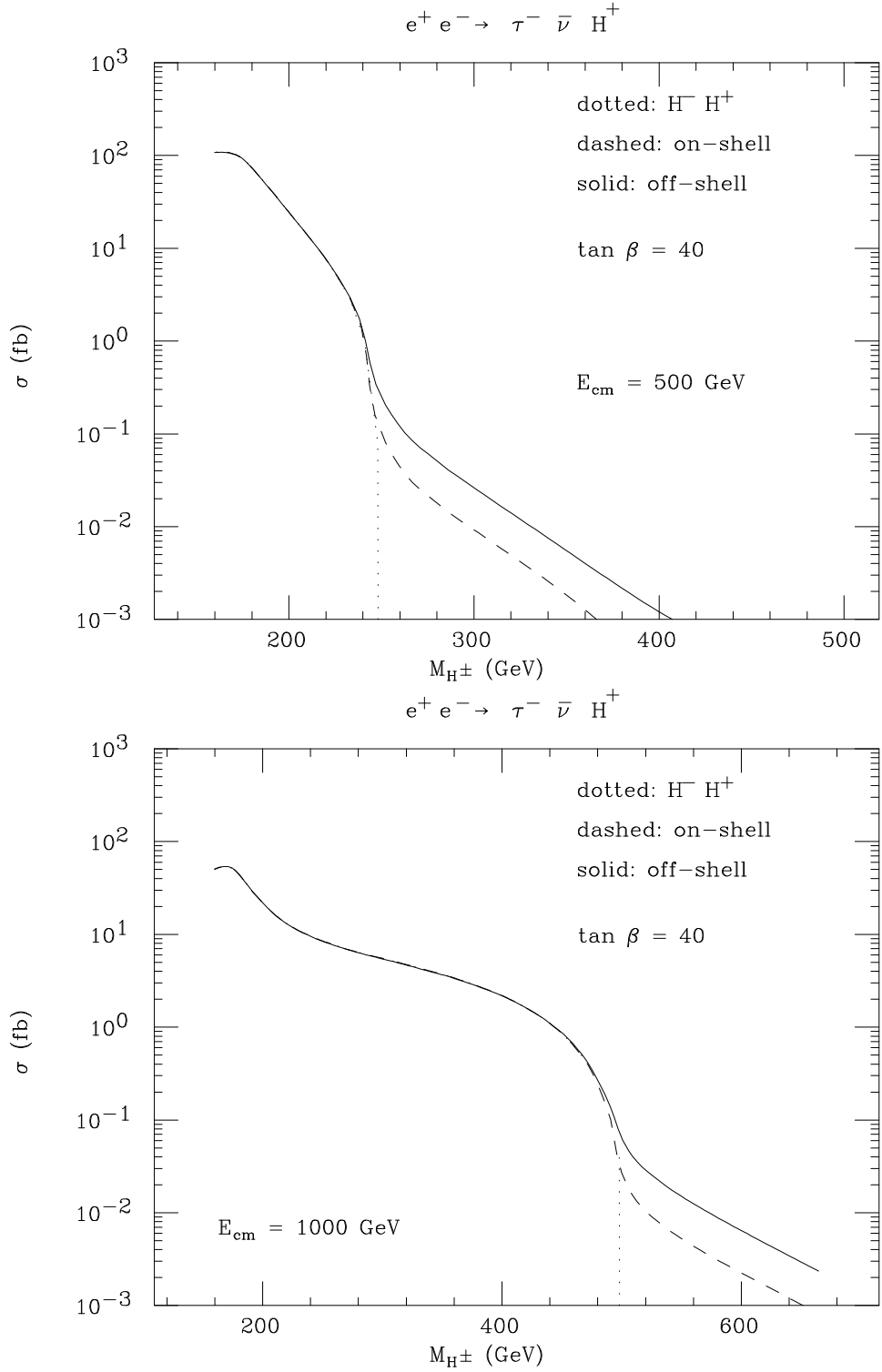

\begin{center}
\epsfig{file=eetnhpm_500_off-shell.ps ,width=10cm,angle=90}
\epsfig{file=eetnhpm_1000_off-shell.ps,width=10cm,angle=90}\\
\end{center}
\vspace{-0.5cm}
\caption{Total cross sections for process (\ref{proc_tau}) with 
the charged Higgs boson being on- and off-shell. No cuts have been 
enforced here. We also show the cross sections corresponding to
$e^+e^-\to H^-H^+$ production times the decay BR for
$H^-\to\tau^-\bar\nu_\tau$.}
\label{fig:off-shell}
\end{figure}

We start our numerical investigation by comparing the LC rates for 
process (\ref{proc_tau}) computed with the charged Higgs boson set on-shell
(also refer to Ref.~\cite{singleHpm}) to those in 
which the latter is allowed to be off-shell. 
The corresponding curves are displayed in
Fig.~\ref{fig:off-shell}. No cuts are enforced here.
At the `threshold' point $M_{H^\pm}\approx
\sqrt s/2$, one may notice that the two curves start departing. The effect
is similar in size and shape at both energies considered. It is due to the
finite width of the charged Higgs boson, which is of several GeV. By rewriting
the Higgs propagator in the off-shell process as 
\begin{equation}\label{prop}
\frac{ p\Dir  + M_{H^\pm}}{p^2-M^2_{H^\pm}+iM_{H^\pm}\Gamma_{H^\pm}}
\left( \frac{\Gamma_{H^\pm}}{\Gamma_{\mathrm{tot}}}\right)^{\frac{1}{2}}
\end{equation}
and taking the limit $\Gamma_{H^\pm} \to 0$, the Breit-Wigner in 
eq.~(\ref{prop}) becomes a representation of the Dirac
delta function $\delta(p^2-M_{H^\pm}^2)$ (apart from a factor
$\pi$) and the on-shell or Narrow
Width Approximation (NWA) is recovered.
For reference, in the same figure, we also show the rates
obtained by using the two-body mode $e^+e^-\to H^-H^+$ followed by
the decay $H^-\to \tau^-\bar\nu_\tau$.

The tails surviving the sharp drop at $M_{H^\pm}\approx
\sqrt s/2$ are due to the diagrams that do not proceed via
$e^+e^-\to H^-H^+\to \tau^-\bar\nu_\tau H^+$ and to the relative interference
between the two sets of graphs. For reference, one should recall that
the top-antitop background is at this stage (including the decay
BRs yielding the four-body final state in (\ref{background})) about 120 and 30 
fb at $\sqrt s=500$ and 1000 GeV, respectively
(at leading order). The $S/B$ ratio is prohibitively large then, to
start with, about 1:600(1:800) at $M_{H^\pm}\approx \sqrt s/2$,
for $\sqrt s=500(1000)$ GeV.

We now proceed our investigation by enforcing some selection cuts.
Like in Ref.~\cite{Battaglia},
the Cambridge jet clustering algorithm \cite{CAMBRIDGE} (see \cite{JETS}
for a comparative review of its properties) 
was enforced to isolate a five jet sample, here with $y_{\rm cut}=0.001$,
wherein the $\tau$-jet was treated on the same footing as the quark-jets.
Similarly, both $\tau$- and quark-jets were required to pass the following
cuts in energy and polar angle (hereafter, $j$ represents a generic jet):
\begin{equation} \label{Ethetacuts}
 E_{j} >  5    ~{\mathrm GeV},\qquad 
 |\cos\theta_{j}| <  0.995.
\end{equation}
However, the former can be distinguished from the latter rather
efficiently, thanks to very different sub-jet distributions (e.g.,
charged hadron multiplicities). Thus, one can apply a sequential $W^\pm$ and
$t$ mass reconstruction only to quark-jets, as follows:
\begin{equation}\label{masscuts}
 |M_{jj}-M_{W^\pm}| <  10    ~{\mathrm GeV},\qquad 
 |M_{jjj}-m_t| <  15    ~{\mathrm GeV}.
\end{equation}
The cut in missing transverse momentum was:
\begin{equation}\label{pTmisscut}
p_T^{\rm miss} >  40    ~{\mathrm GeV}.
\end{equation}
Finally, a very useful variable in distinguishing between signal
and background is a transverse mass, $M_T$, constructed from the 
visible $\tau$-jet and the missing transverse momentum, i.e.,
\begin{equation}\label{MT} 
M_T =\sqrt{2 p_T^{\tau}
{p}_T^{\mathrm{miss}} (1 - \cos\Delta\phi)},
\end{equation}
where $\Delta\phi$ is the relative azimuthal angle.
In the case of the signal, the 
$\tau$-jets are heavily boosted relatively to the case of the background, 
as the charged Higgs masses considered here are much heavier 
than $M_{W^\pm}$. 
By imposing  
\begin{equation}\label{MTcut}
M_T >M_{W^\pm}\approx {\mathrm 80~GeV},
\end{equation}
the background is severely rejected while most of the signal
survives.

Now, recall that distributions of $\pi^\pm$ tracks coming from 
$\tau$'s are sensitive to the polarisation state of the 
$\tau$-lepton and that in turn the spin/helicity 
states of $\tau$'s coming from $H^\pm$ scalars and $W^\pm$ gauge
bosons are different \cite{tau_polarisation,Was,charged}.
In fact, the exploitation of this fact has already been proved
to be very effective in the detection of $H^+\to \tau^+\nu_\tau$
signals in hadron-hadron collisions 
\cite{mono,tau_exp,LesHouches2001}\footnote{For the 
effects of $\tau$-polarisation 
in the neutral Higgs sector, see \cite{tautau}.}.

Basically, the key feature relevant to our purposes is the correlation
between the polarisation state of the decaying boson
and the energy sharing among the emerging pions.
In fact, it is to be noted that the spin
state of $\tau$'s coming from $H^\pm$- and $W^\pm$-boson
decays are opposite: i.e., $H^- \rightarrow \tau^-_R \bar\nu_R$ and $
H^+ \rightarrow \tau^+_L \nu_L$ whereas
$W^- \rightarrow \tau^-_L \bar\nu_R$ and $W^+ \rightarrow \tau^+_R \nu_L$
(neglecting leptonic mass effects, as we did here).
Ultimately, this leads to a significantly harder momentum
distribution of charged pions from $\tau$-decays for the 
$H^\pm$-signal compared to the $W^\pm$-background, which can then be exploited
to increase  $S/B$. This is true
for the case of one-prong decays into both $\pi^\pm$'s and longitudinal
vector mesons, while the transverse component of the latter dilutes
the effect and must be somehow eliminated. This can be done inclusively, 
i.e., without having to identify the individual mesonic component 
of the one-prong hadronic topology. In doing so, we will 
closely follow Ref.~\cite{dptau}.  

\begin{figure}
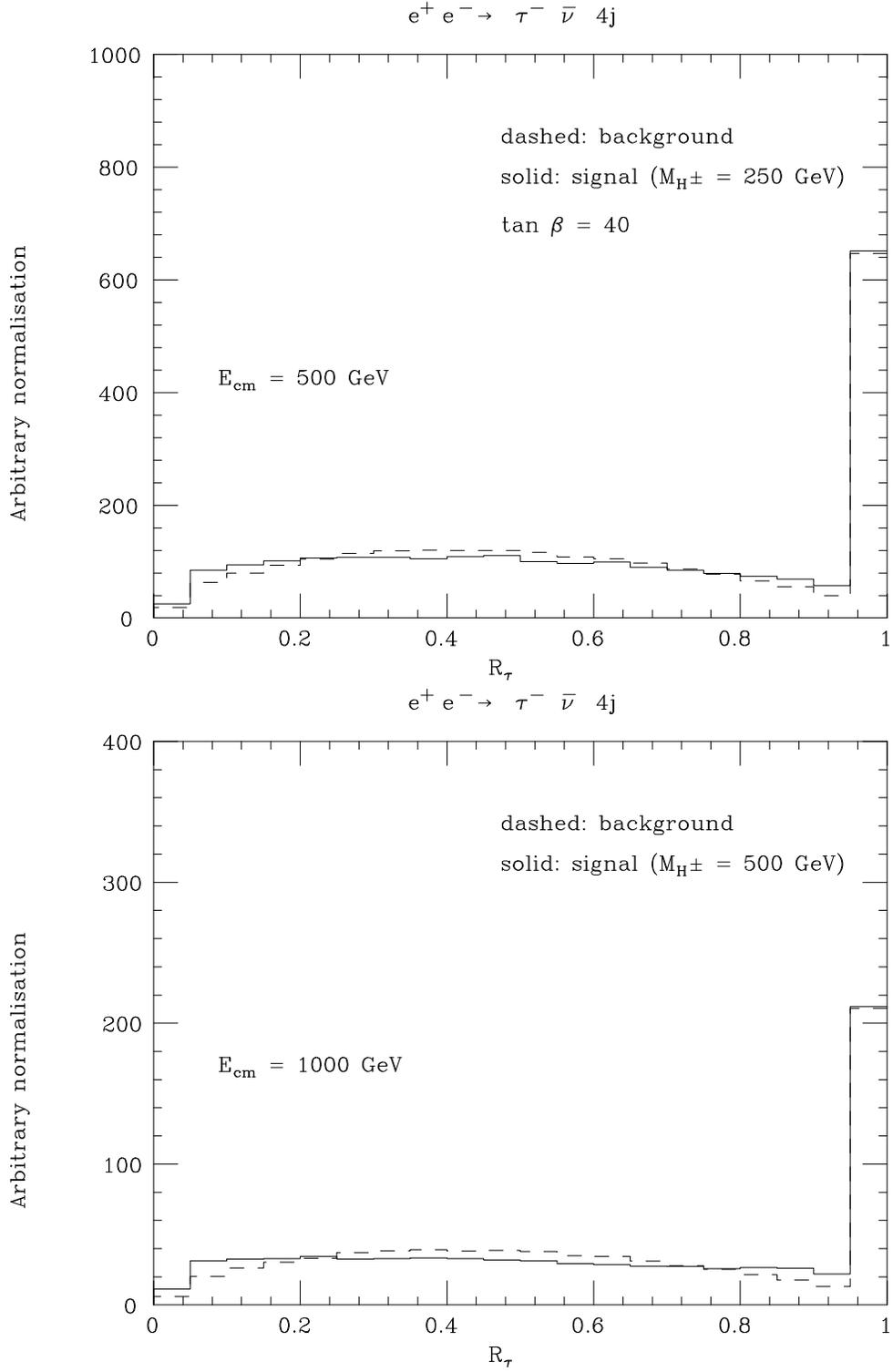

\begin{center}
\epsfig{file=eetnhpm_500_polar.ps ,width=10cm,angle=90}
\epsfig{file=eetnhpm_1000_polar.ps,width=10cm,angle=90}\\
\end{center}
\vspace{-0.5cm}
\caption{Differential distribution in the quantity defined
in (\ref{fraccut}) for processes (\ref{signal}) and (\ref{background}). 
No cuts have been enforced here. Histograms are 0.05 units wide.}
\label{fig:Rtau}
\end{figure}

The mentioned transverse components of the signal as well as those 
of the background can adequately be suppressed
by requiring that 80\% of the $\tau$-jet (transverse) 
energy is carried away by 
the $\pi^\pm$'s, i.e.: 
\begin{equation}\label{fraccut}
R_\tau=\frac{p^{\pi^\pm}}{p_T^{\tau}}> 0.8.
\end{equation}
These arguments certainly applies in the region $M_{H^\pm}\lsim \sqrt s/2$,
where the production cross section of process (\ref{proc_tau}) is
dominated by the two-body mode $e^+e^-\to H^-H^+$ followed by
the decay $H^-\to \tau^-\bar\nu_\tau$. Here, the
enforcement of the constraint in (\ref{fraccut}) 
reduces significantly the background
while costing very little to the signal. However, we are mostly
interested in the complementary mass interval, $M_{H^\pm}\gsim \sqrt s/2$,
where pair production becomes irrelevant. Here, we have verified that
the enforcement of the cut in (\ref{fraccut}) is harmless with
respect to the effects on the $S/B$ rates, as one can appreciate from 
Fig.~\ref{fig:Rtau}, showing that the $R_\tau$ distributions
for signal and background are basically identical. Therefore, we maintained
the requirement (\ref{fraccut})
throughout the  $M_{H^\pm}$ range considered here, for sake of
consistency.

\begin{figure}
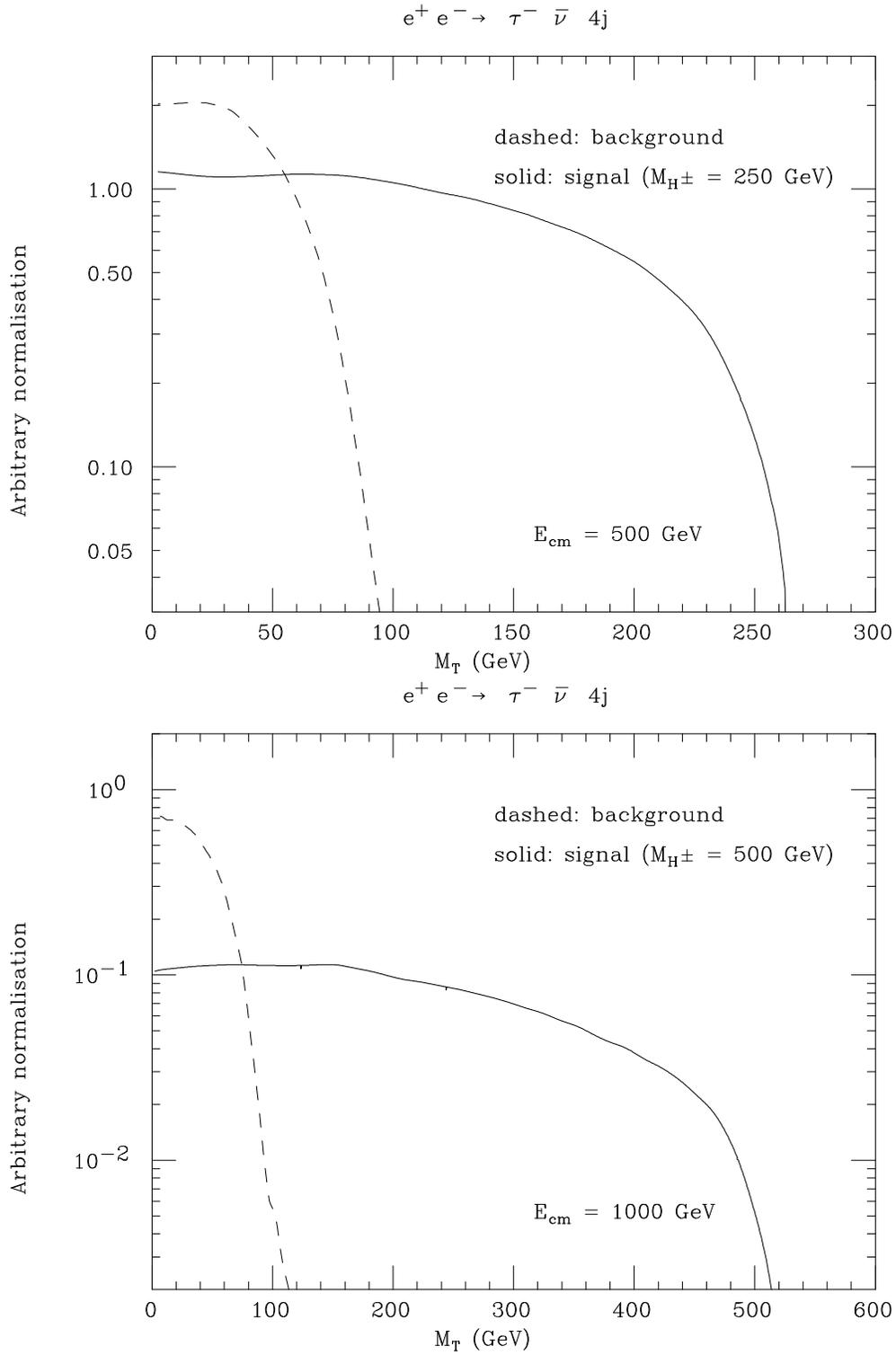

\begin{center}
\epsfig{file=eetnhpm_500_MT.ps ,width=10cm,angle=90}
\epsfig{file=eetnhpm_1000_MT.ps,width=10cm,angle=90}\\
\end{center}
\vspace{-0.5cm}
\caption{Differential distribution in the transverse mass (\ref{MT})
for processes (\ref{signal}) and (\ref{background}). No cuts have been 
enforced here.}
\label{fig:MT}
\end{figure}

A vigorous reduction of the background rates can however 
be obtained by enforcing all the other kinematic cuts described above:
(\ref{Ethetacuts})--(\ref{MTcut}). As intimated,  
Fig.~\ref{fig:MT} illustrates the strong impact of the constraint in
transverse mass, by comparing the shape of the signal and background
before the kinematic selection. The signal distributions are obtained
at the points $M_{H^\pm}\approx \sqrt s/2$, at both centre-of-mass
(CM) energies.

\begin{figure}
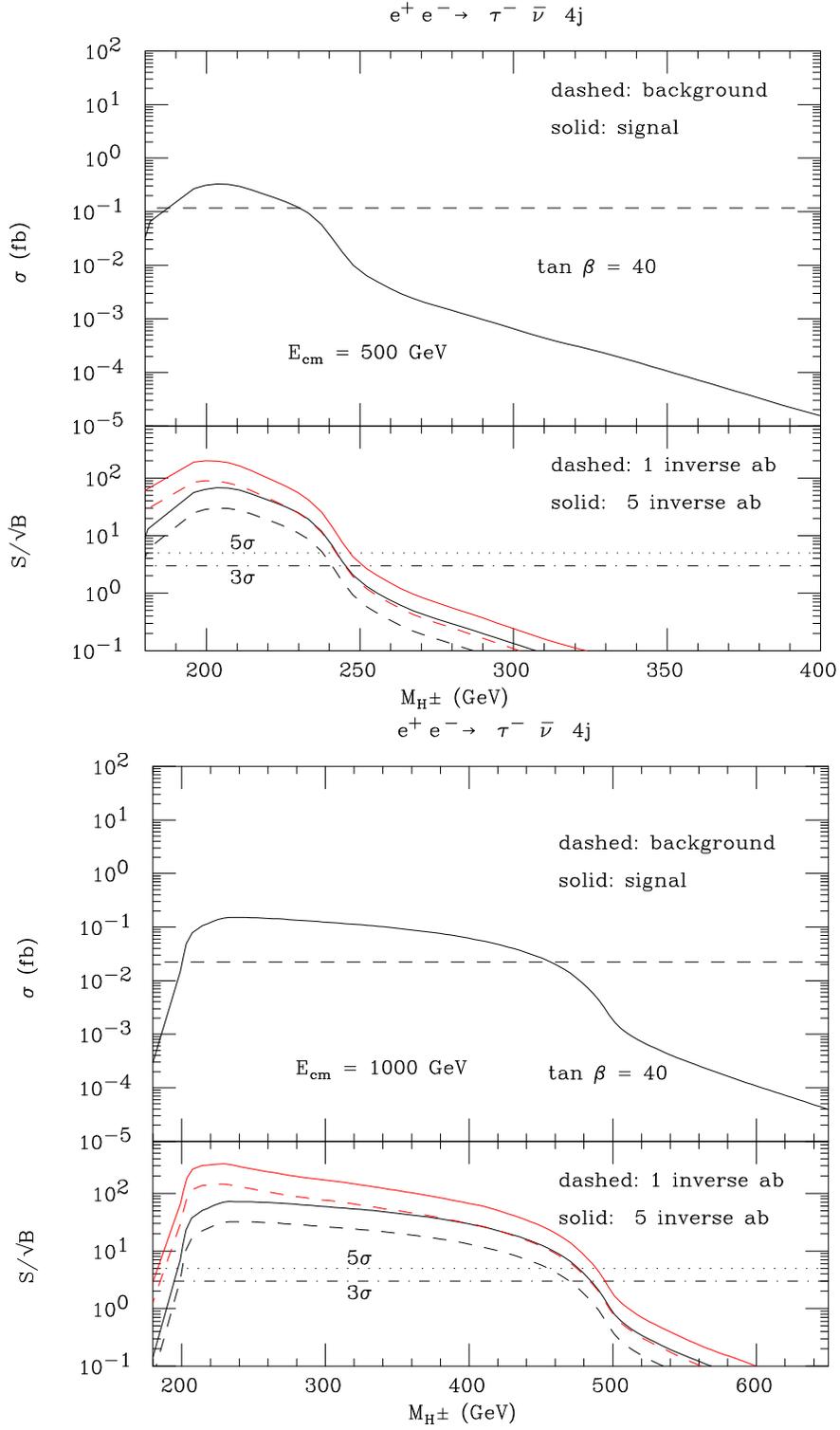

\begin{center}
\epsfig{file=eetnhpm_500_cuts.ps ,width=10cm,angle=90}
\epsfig{file=eetnhpm_1000_cuts.ps,width=10cm,angle=90}\\
\end{center}
\vspace{-0.5cm}
\caption{(Top) Total cross sections for processes 
(\ref{signal}) and (\ref{background}) yielding the signature (\ref{signature}),
after the kinematic cuts in (\ref{Ethetacuts})--(\ref{fraccut}), including 
all decay BRs. 
(Bottom) Statistical significances of the signal
for two values of integrated
luminosity (the $3\sigma$ and $5\sigma$ `evidence' and `discovery' 
threshold are also given) after the kinematic
cuts in (\ref{Ethetacuts})--(\ref{fraccut}) (in black) and the
additional one in (\ref{M4jcut}) (in red).}
\label{fig:cuts}
\end{figure}

The upper plots in Fig.~\ref{fig:cuts} present the signal rates after the 
full kinematic selection has been enforced. (The background 
cross section is constant with $M_{H^\pm}$ as the above 
cuts do not depend on this
parameter.)  In the lower plots we display the significances
(in black) of the signal rates, after 1 and 5 ab$^{-1}$ of accumulated
luminosity, $\cal L$. 
It is clear that at this point neither evidence ($\gsim3\sigma$)
nor discovery ($\gsim5\sigma$) of charged Higgs bosons is possible
in the region $M_{H^\pm}\gsim\sqrt s/2$, whereas for
$M_{H^\pm}\lsim\sqrt s/2$ the signals should be comfortably observed.

\begin{figure}
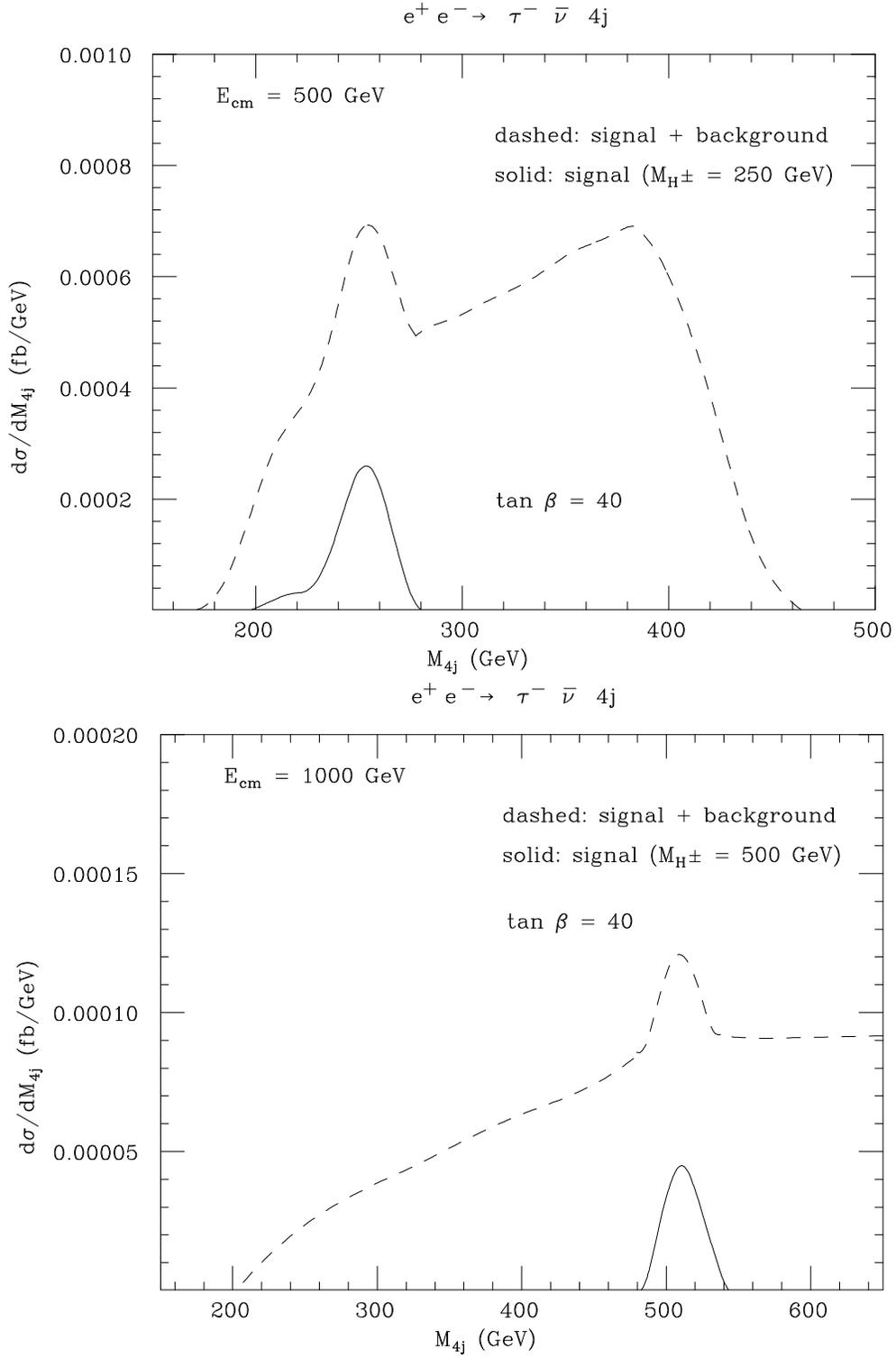

\begin{center}
\epsfig{file=eetnhpm_500_mass.ps ,width=10cm,angle=90}
\epsfig{file=eetnhpm_1000_mass.ps,width=10cm,angle=90}\\
\end{center}
\vspace{-0.5cm}
\caption{Differential distribution in the invariant mass of the four quark-jets
recoiling against the $\tau$-jet
for the sum of processes (\ref{signal}) and (\ref{background}) 
and for the former separately yielding the signature (\ref{signature}),
after the
kinematic cuts in (\ref{Ethetacuts})--(\ref{fraccut}), including 
all decay BRs.}
\label{fig:mass}
\end{figure}

At this point, one should recall that the charged Higgs bosons in the final
state decay into visible objects, i.e., four quark-jets. Besides, 
recalling that the latter can efficiently be distinguished from
$\tau$-jets, one can look at the invariant mass of this multi-jet system.
Fig.~\ref{fig:mass} shows this quantity. For the signal, it represents
the reconstructed resonance of the charged Higgs boson, that we have 
originally generated at the point $M_{H^\pm}\approx \sqrt s/2$, for both CM
energies. For the background, it correspond to a non-resonant kinematic
distribution. The width of the signal spectra is dominated by
detector smearing effects and suggests that a further selection
criterium can be enforced to enhance the $S/B$ rates, e.g.:
\begin{equation}\label{M4jcut}
|M_{4j}-M_{H^\pm}|<35~{\rm GeV}. 
\end{equation}
The value of $M_{H^\pm}$ entering eq.~(\ref{M4jcut}) would be the central 
or fitted mass resonance of the region in $M_{4j}$ were an excess of the form 
seen in Fig.~\ref{fig:mass} will be established.

The two red lines in Fig.~\ref{fig:cuts}~show the significances
of the charged Higgs boson signals in presence of the constraint
in (\ref{M4jcut}), alongside those in (\ref{Ethetacuts})--(\ref{fraccut}).
By comparing these curves with the shape of the two-body $H^-H^+$ cross
section in Fig.~\ref{fig:off-shell}, one should expect to extend the reach 
in $M_{H^\pm}$ obtained from pair production of charged Higgs bosons and 
decays by about 20--30 GeV or so around and above the $M_{H^\pm}= \sqrt s/2$
point, at both CM energies, thanks to the contribution of
single $H^\pm$ production. Typical signal rates at $\tan\beta=40$ 
in the threshold
region would be 6(30) events at $\sqrt s=500$ GeV and 2(10)
at $\sqrt s=1000$ GeV, in correspondence of $\cal L=$ 1(5) ab$^{-1}$.
Furthermore, recall that in the high $\tan\beta$ region, where process
(\ref{proc_tau}) is of relevance, its production rates approximately
scale like $\tan\beta^2$, so that the higher this parameter the better
the chances of isolating the signal discussed here. Finally,
in our estimates so far, 
we have excluded the efficiency $\epsilon^2_b$ of tagging the two $b$-jets
in the final state. According to Ref.~\cite{Battaglia}, the single
$b$-tag efficiency is expected to be close to the value $\epsilon_b=90\%$, 
so that our main 
conclusions should remain unchanged.

We regard our findings as rather encouraging, especially considering
the initial value of the $S/B$ rates. Indeed, better selection procedures
than those outlined here could be devised
(see, e.g., \cite{Battaglia}), to improve further the discovery
reach of heavy charged Higgs bosons, with masses $M_{H^\pm}\gsim \sqrt s/2$,
in a general 2HDM (including the MSSM).
A realistic analysis exploiting  more sophisticated simulations, 
based on the HERWIG Monte Carlo event generator \cite{HERWIG,me}
interfaced to the TAUOLA package \cite{TAUOLA} (see also \cite{Worek}) 
for polarised $\tau$-decays,  will be available
in the near future \cite{Preparation}.

\clearpage

\clearpage

\clearpage

\clearpage

\clearpage

\begin{thebibliography}{99}

{\small

\bibitem{eebiblio} K.~Abe \etal, [The ACFA Linear Collider Working Group],
hep-ph/0109166
and references therein;
T.~Abe \etal, [The American Linear Collider Working Group], 
hep-ex/0106055; hep-ex/0106056; hep-ex/0106057 and hep-ex/0106058
and references therein; 
J.A. Aguilar-Saavedra \etal, [The 
ECFA/DESY LC Physics Working Group],  hep-ph/0106315;
G. Guignard (editor), [The CLIC Study Team], preprint CERN-2000-008.


\bibitem{singleHpm} 
A.~Djouadi, J.~Kalinowski and P.~M.~Zerwas, Z.\ Phys.\ C{54}
(1992) 255; S. Kanemura, S. Moretti and K. Odagiri, 
JHEP 02 (2001) 011; B.A. 
Kniehl, F. Madricardo and M. Steinhauser, hep-ph/0205312.

\bibitem{proceedings} S. Kanemura, S. Moretti and K. Odagiri,
hep-ph/0101354.

\bibitem{Lorenzo} A.~Gutierrez-Rodriguez and O.A.~Sampayo,
                     {hep-ph/9911361}.

\bibitem{Logan} H.E. Logan and S. Su,  hep-ph/0203270; hep-ph/0206135.

\bibitem{WH} S.H.~Zhu,  hep-ph/{9901221};
S.~Kanemura, Eur. Phys. J. {C17} ({2000}) {473};
A.~Arhrib, M.~Capdequi-Peyran\`ere, W.~Hollik and
                     G.~Moultaka, Nucl. Phys. {B581} ({2000}) {34}.

\bibitem{Higgs} 
A. Djouadi, J. Kalinowski and P.M. Zerwas, in proceedings of the 
Workshop `$e^+e^-$ collisions at 500 GeV: The Physics Potential', part A,
preprint DESY 92-123A, August 1992;
Z.\  Phys.\ C57 (1993) 569.   

\bibitem{eeHpHm} 
S. Komamiya, Phys. Rev. D38 (1988) 2158.

\bibitem{egamma} S. Kanemura, S. Moretti and K. Odagiri, Eur. Phys.
J. C22 (2001) 401.

\bibitem{2gamma} S. Kanemura, S. Moretti and K. Odagiri, 
in preparation.

\bibitem{LEPTRE} For a review, see, e.g.:
M. Antonelli and S. Moretti, hep-ph/0106332 (and references therein).

\bibitem{BRs} S. Moretti and W.J. Stirling,
Phys. Lett. B347 (1995) 291; Erratum, ibidem B366 (1996) 451. 

\bibitem{tt} A. Ballestrero, E. Maina and S. Moretti,
Phys. Lett. B333 (1994) 434; Z. Phys. C72 (1996) 71.

\bibitem{bbWW} A. Ballestrero, E. Maina and S. Moretti,
Phys. Lett. B335 (1994) 460; S. Moretti, Z. Phys. C73 (1997) 653. 

\bibitem{bbZZ} S. Moretti, Phys. Rev. D52 (1995) 6316.

\bibitem{bbCC} S. Moretti and K. Odagiri, Eur. Phys. J. C1 (1998) 633.

\bibitem{W4j} S. Moretti, Z. Phys. C75 (1997) 465; 
Eur. Phys. J. C9 (1999) 229.  

\bibitem{VEGAS} G.P. Lepage, {Jour. Comp. Phys.} {27} (1978) 192.

\bibitem{Metro} H. Kharraziha and S. Moretti, 
       Comp. Phys. Comm. 127 (2000) 242;
 Erratum, ibidem B134 (2001) 136. 

\bibitem{Battaglia} 
 M. Battaglia, A. Ferrari, A. Kiiskinen and T. Maki, hep-ex/0112015.

\bibitem{CAMBRIDGE} G. Leder, S. Moretti and B.R. Webber,
JHEP {08} (1997) 001.

\bibitem{JETS} 
L. L\"onnblad, S. Moretti  and T. Sj\"ostrand,
       JHEP {08} (1998) 001.

\bibitem{tau_polarisation}
B.K. Bullock, K. Hagiwara and A.D. Martin,
Nucl. Phys. B395 (1993) 499.

\bibitem{Was}
T. Pierzchala, E. Richter-Was, Z. Was and M. Worek,
Acta Phys. Polon. B32 (2001) 1277.

\bibitem{charged} S. Raychaudhuri and D.P. Roy, 
Phys. Rev. D53 (1996) 4902; D.P. Roy, Phys. Lett. B277 (1992) 183; 
Phys. Lett. B459 (1999) 607. 

\bibitem{mono} M. Guchait and S. Moretti, JHEP 01 (2002) 001.

\bibitem{tau_exp} 
R. Kinnunen, CMS-NOTE-2000/045;
K.A. Assamagan and Y. Coadou,
Acta Phys. Polon. B33 (2002) 707;
ATL-PHYS-2001-031; 
K.A. Assamagan, A. Djouadi, M. Drees,
M. Guchait, R. Kinnunen, J.L. Kneur, D.J. Miller, S. Moretti, K. Odagiri
and D.P. Roy, contributed to 
the `The Higgs Working Group: Summary Report' of the
Workshop `Physics at TeV Colliders',
Les Houches, France, 8-18 June 1999,
{hep-ph/0002258}.
        
\bibitem{LesHouches2001} 
K.A. Assamagan, Y. Coadou and A. Deandrea, hep-ph/0203121; 
K.A. Assamagan, M. Bisset, Y. Coadou, A.K. Datta, 
A. Deandrea,
A. Djouadi, M. Guchait, Y. Mambrini, F. Moortgat and
S. Moretti, contributed to the `The Higgs Working Group: Summary Report' of the
Workshop `Physics at TeV Colliders',
Les Houches, France, 21 May-1 June 2001,
{hep-ph/0203056}.

\bibitem{tautau} S. Moretti and D.P. Roy, 
hep-ph/0206206.

\bibitem{dptau}
S. Raychaudhuri and D.P. Roy, Phys. Rev. {D52} (1995) 1556.

\bibitem{HERWIG}
G. Marchesini, B.R. Webber,  G. Abbiendi, I.G. Knowles, M.H. Seymour
and L. Stanco, Comput.\ Phys.\ Commun.\ {67} (1992) 465;
G. Corcella, I.G. Knowles, G. Marchesini,  
S. Moretti, K. Odagiri, P. Richardson, M.H. Seymour and  B.R. Webber,
 {hep-ph/9912396}; JHEP {01} (2001) 010;
 {hep-ph/0107071}; {hep-ph/0201201}.

\bibitem{me} S. Moretti, K. Odagiri, P. Richardson, M.H. Seymour
and B.R. Webber, JHEP 04 (2002) 028; S. Moretti,  {hep-ph/0205105}.

\bibitem{TAUOLA}
S. Jadach, Z. Was, R. Decker and J.H. K\"uhn,
Comput. Phys. Commun. 76 (1993) 361; 
M. Jezabek, Z. Was, S. Jadach and J.H. K\"uhn,
Comput. Phys. Commun. 70 (1992) 69; 
S. Jadach, J.H. K\"uhn and Z. Was,
Comput. Phys. Commun. 64 (1990) 275. 

\bibitem{Worek} 
M. Worek, Acta Phys. Polon. B32 (2001) 3803.

\bibitem{Preparation} S. Moretti,
in preparation.

}

\end{thebibliography}
\end{document}